\documentclass[]{spie}  

\usepackage{siunitx}
\usepackage{amsmath,amsfonts,amssymb}
\usepackage{graphicx}
\usepackage[colorlinks=true, allcolors=blue]{hyperref}
\usepackage{fancyhdr}

\title{Review of Radio Frequency Interference and Potential Impacts on the CMB-S4 Cosmic Microwave Background Survey}

\author[a]{Darcy R. Barron}
\author[b,c,d]{Amy N. Bender}
\author[a]{Ian E. Birdwell}
\author[b,c,d]{John E. Carlstrom}
\author[e]{Jacques Delabrouille}
\author[f]{Sam Guns}
\author[g]{John Kovac}
\author[h]{Charles R. Lawrence}
\author[g]{Scott Paine}
\author[i]{Nathan Whitehorn}

\affil[a]{Department of Physics and Astronomy, University of New Mexico, Albuquerque, NM, 87131, USA}
\affil[b]{High-Energy Physics Division, Argonne National Laboratory, 9700 South Cass Avenue., Lemont, IL, 60439, USA}
\affil[c]{Department of Astronomy and Astrophysics, University of Chicago, 5640 South Ellis Avenue, Chicago, IL, 60637, USA}
\affil[d]{Kavli Institute for Cosmological Physics, University of Chicago, 5640 South Ellis Avenue, Chicago, IL, 60637, USA}
\affil[e]{CNRS-UCB International Research Laboratory, Centre Pierre Binétruy, IRL2007, CPB-IN2P3, Berkeley, CA, 94720, USA}
\affil[f]{Department of Physics, University of California, Berkeley, CA, 94720, USA}
\affil[g]{Center for Astrophysics \textbar\ Harvard \& Smithsonian, Cambridge, MA, USA}
\affil[h]{Jet Propulsion Laboratory, California Institute of Technology, 4800 Oak Grove Drive, Pasadena, CA 91109, USA}
\affil[i]{Department of Physics and Astronomy, Michigan State University, East Lansing, MI 48824, USA}

\authorinfo{Further author information: (Send correspondence to D.R.B) \\D.R.B.: E-mail: dbarron2@unm.edu}

\fancypagestyle{FirstPage}{
\lfoot{Copyright 2022
Society of PhotoOptical Instrumentation Engineers. One print or electronic copy may bemade for personal use only. Systematic reproduction and distribution, duplication of any material in this paperfor a fee or for commercial purposes, or modifi cation of the content of the paper are prohibited.} 
}

\begin{document} 
	\maketitle
	
	\begin{abstract}
		CMB-S4 will map the cosmic microwave background to unprecedented precision, while simultaneously surveying the millimeter-wave time-domain sky, in order to advance our understanding of cosmology and the universe. CMB-S4 will observe from two sites, the South Pole and the Atacama Desert of Chile. A combination of small- and large-aperture telescopes with hundreds of thousands of polarization-sensitive detectors will observe in several frequency bands from 20--300\,GHz, surveying more than 50\% of the sky to arcminute resolution with unprecedented sensitivity. CMB-S4 seeks to make a dramatic leap in sensitivity while observing across a broad range of largely unprotected spectrum which is increasingly being utilized for terrestrial and satellite transmissions. Fundamental aspects of CMB instrument technology leave them vulnerable to radio frequency interference (RFI) across a wide range of frequencies, including frequencies outside of their observing bands. Ground-based CMB instruments achieve their extraordinary sensitivities by deploying large focal planes of superconducting bolometers to extremely dry, high-altitude sites, with large fractional bandwidths, wide fields of view, and years of integration time. Suitable observing sites have historically offered significant protection from RFI, both naturally through their extremely remote locations as well as through restrictions on local emissions. Since the coupling mechanisms are complex, “safe” levels or frequencies of emission that would not interfere with CMB measurements cannot always be determined through straightforward calculations. We discuss models of interference for various types of RFI relevant to CMB-S4, mitigation strategies, and the potential impacts on survey sensitivity.
	\end{abstract}
	
	
	
	\section{INTRODUCTION} \label{sec:intro}  
	\thispagestyle{FirstPage}

	The cosmic microwave background (CMB) is relic radiation emitted in the early universe when electrons first combined with light nuclei to form neutral atoms. The CMB is one of the key observables for understanding the history of the universe, its matter and energy content, the formation of structures observable today,
	as well as for probing fundamental particles and interactions between them beyond what can be achieved in human-made particle accelerators.
	Modern ground-based instruments utilize large arrays of superconducting transition edge sensor bolometers operating at background-limited sensitivities in order to achieve precision measurements of the CMB, including its faint polarization signal.
	This technique is extremely powerful, but also quite vulnerable to radio frequency interference (RFI).
	Each bolometer is exceptionally sensitive across a broad fractional bandwidth defined by an on-chip superconducting circuit.
	At extremely dry, high altitude observatories, these bolometers typically measure only a few picowatts of power integrated across 20 GHz of bandwidth, coming from residual atmospheric emission.
	
	
	The success of recent CMB instruments and the potential for even greater discoveries has led to the development of CMB-S4. This next-generation ground-based CMB instrument has had strong community support leading up to its establishment as a DOE project and a formal scientific collaboration, and was recently recommended by the Astro2020 decadal survey in astronomy\cite{astro2020}.
	Precision measurements of the CMB must measure across a broad frequency range around the CMB spectral peak in order to identify and separate emission from non-cosmological sources, including emission originating from the Galactic interstellar medium.
	CMB-S4 will deploy hundreds of thousands of superconducting detectors in frequency bands from 20--300 GHz in order to reach its theoretically motivated measurement targets in cosmology and particle physics.
	To achieve its science goals, CMB-S4 will deploy telescopes to two unique and complementary established CMB observing sites: the Dark Sector at the Amundsen-Scott South Pole Station, and Cerro Toco within the Chajnantor Science Reserve in the high Atacama desert of northern Chile.
	
	The survey for CMB-S4 necessarily uses unprotected spectrum allocated to active services. Prior astronomical surveys using the same unprotected spectrum have only been made possible through geographic isolation of observatories in these radio quiet zones in extremely remote locations.  The rapid development of large commercial satellite constellations for telecommunications is now placing such geographic protection under threat.  The continued feasibility of leading-edge science including CMB observations may come to depend on protection via explicit geographic quiet zones extending into LEO.
	Because of their geographic isolation, both of these observatory sites have thus far enjoyed relative freedom from commercial or public radio emissions.  Instead, the main interference sources have been local transmitters related to site operations, and have been managed by explicit coordination or ad hoc identification and mitigation of RFI.  The rapid emergence of large commercial satellite constellations will change this, and understanding their impact and developing mitigation strategies is a matter of urgency.  The ideal would be coexistence whereby expanded communications bandwidth made available by the satellite constellations can be used to support the scientific mission without degrading or disabling it. 
	
	In these proceedings, we give a review of RFI in modern cosmic microwave background instruments, including mitigation techniques and their limitations, with the goal of motivating concerns about the potential for significant impacts from RFI on the upcoming CMB-S4 cosmic microwave background survey.
	In Section \ref{sec:intro}, we describe the history of CMB measurements and the motivation for the next-generation experiment CMB-S4, as well as background on radio spectrum protections for radio astronomy. In Section \ref{sec:instrumentdesign} we give an overview of modern CMB instrument design, describing the enabling technologies including broadband bolometric detectors for ultra-sensitive CMB instruments. In Section \ref{sec:coupling} we discuss the different modes of RFI coupling to a CMB instrument and instrument-level mitigation techniques, including indirect coupling, coupling to out-of-band RFI, and coupling to in-band RFI. Section \ref{sec:rfi_pole} gives historical examples of harmful interference from RFI impacting CMB instruments at the South Pole's Dark Sector, and steps taken for mitigation. Section \ref{sec:satelliterfi} focuses on the special case of RFI originating from communications satellites, which can couple directly to the instrument's astronomical signal path. Section \ref{sec:analysis_mitigation} describes the impact of RFI on observations and mitigation measures taken in analysis to remove spurious signals from the final cosmological survey.
	
	\subsection{Overview of Cosmic Microwave Background Measurements} 
	
	The cosmic microwave background was originally emitted with a near-blackbody spectrum at a glowing-hot temperature of $\simeq$\,3000\,K, comparable to the surface temperature of a star such as the Sun. 
	The CMB photons have been cooled down adiabatically by the expansion of the universe to about 2.7\,K today---a loss of energy by a factor about 1100 for each CMB photon. Although their present frequency distribution peaks at about 160\,GHz, which is a preferred frequency for sensitive CMB observations, CMB photons can be detected over a large range of frequencies extending from radio to sub-millimeter wavelengths. The first detection of the CMB was made serendipitously by Bell
	 Labs physicists Arno Penzias and Robert Wilson in 1964 while testing a radiocommunication antenna, at a frequency of 4080\,MHz\cite{1965ApJ...142..419P}. This detection had profound consequences on the field of cosmology, strongly supporting the model of an expanding universe starting with a hot Big-Bang, while ruling out all of the alternatives that were considered at the time\cite{1965ApJ...142..414D}.
	
	The extraordinary homogeneity of the temperature of the CMB over the sky, and its near-blackbody spectrum\cite{1994ApJ...420..439M}, testify of a very homogeneous, dense and hot universe at early times. However, the present existence of cosmic structures requires the existence of initial small perturbations of the matter and radiation density, from which structures grow by the action of gravity. These density perturbations leave, in the CMB temperature, an imprint in the form of small temperature anisotropies, detected for the first time in 1992 by the DMR instrument aboard the COBE satellite\cite{1992ApJ...396L...1S}.
	
	In the next decades, the observation of CMB temperature anisotropies and polarization patterns has been further successful in constraining the properties of our Universe as a whole, such as its global geometry (with spatially flat hypersurfaces and expansion over time), its age, and its matter and energy content. This progress has been made possible, over the years, by a wide range of observations from the ground\cite{2002Natur.420..772K,2009ApJ...694.1200R,2009ApJ...705..978B}, from stratospheric balloons\cite{2000ApJ...545L...5H,2002ApJ...564..559D,2003A&A...399L..19B}, and from space\cite{2013ApJS..208...20B,2011A&A...536A...1P}, culminating with the recent results of the Planck space mission\cite{2020A&A...641A...1P}.
	
	At this point, however, still only a fraction of the cosmological information available in the CMB has been exploited. More observations 
	will be required to harvest, in particular, the information encoded in the polarization of the CMB at large angular scale. CMB polarization can be decomposed in a linear superposition of even-parity modes (called the E-modes), and odd-parity modes (the B-modes)\cite{1997PhRvD..55.7368K,1997PhRvL..78.2054S}. B-modes, which cannot be generated in the early universe by density inhomogeneities alone, are a tracer of the existence of primordial gravitational waves. As such, they are a probe of an epoch in which such gravitational waves might have been generated, the epoch of cosmic inflation, which has been postulated as an explanation for the generation of the seeds of structures, and of the observed spatial flatness of the universe. While an appealing theory, compatible with existing observations, cosmic inflation still lacks direct observational confirmation. The detection of primordial CMB B-modes would be a smoking gun in this respect, constraining physical processes at work at energies $10^{12}$ times larger than those available on the largest human-made particle accelerator to date, the LHC\cite{2016ARA&A..54..227K}.
	
\begin{figure} [ht]
		\begin{center}
			\begin{tabular}{cc} 
				\includegraphics[height=6cm]{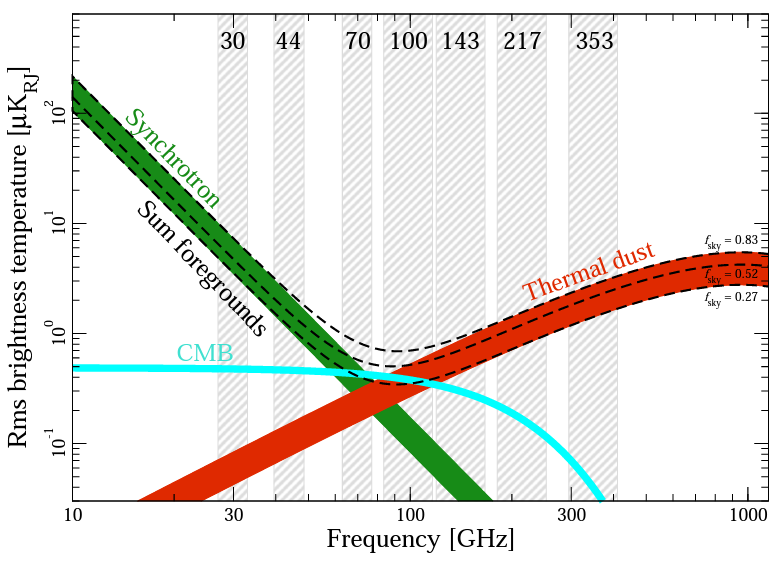}
				\includegraphics[height=6.2cm]{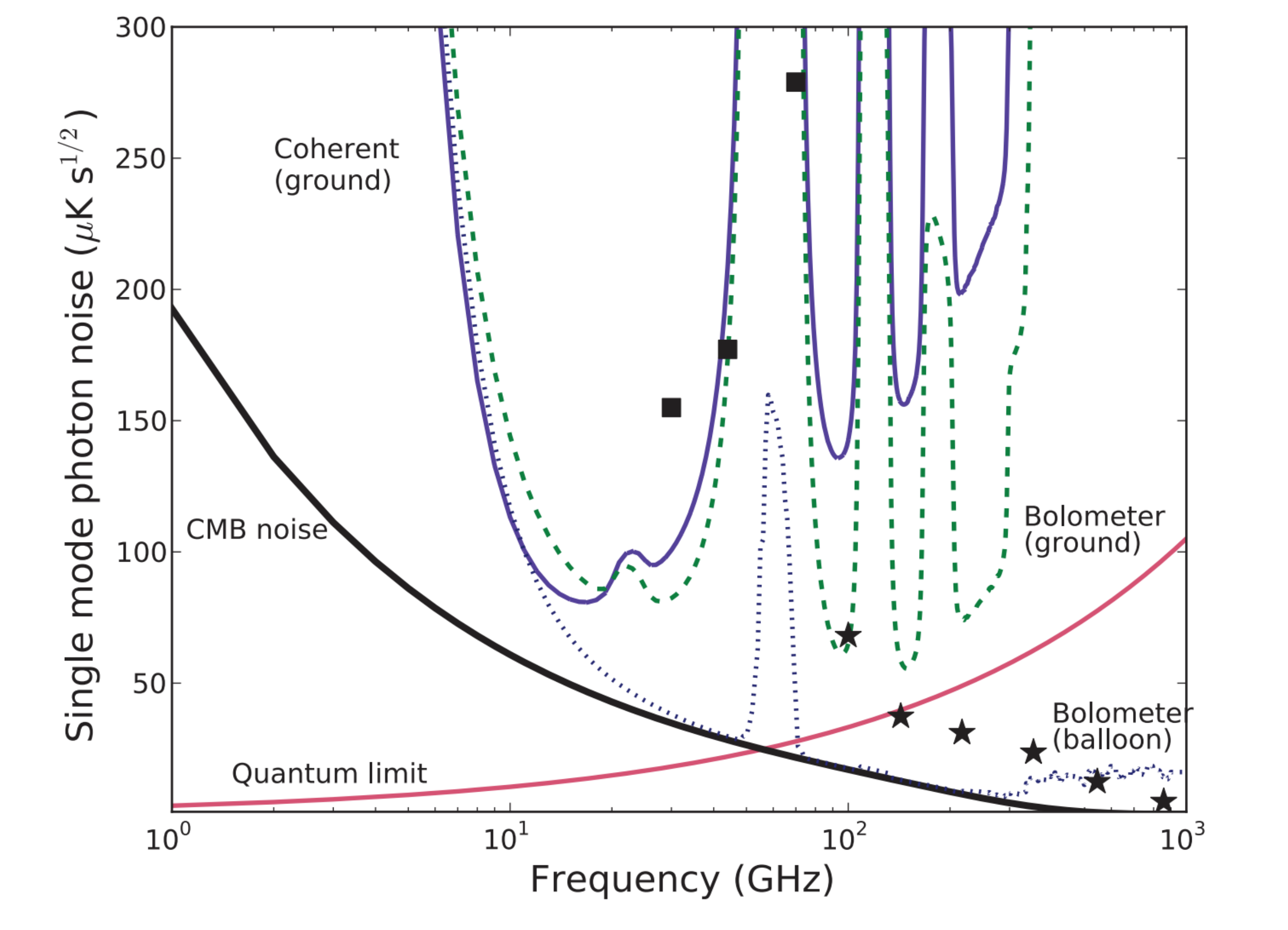}
			\end{tabular}
		\end{center}
		\caption[cmbspectrum] 
		{ \label{fig:cmb_spectrum} 
			(Left) The CMB polarization frequency spectrum (teal) is shown along with other bright astrophysical components. The measurement bands for the Planck satellite are also shown (grey vertical bands). Figure from Ref. \citenum{planck_collaboration_planck_2014}. (Right) A comparison of modeled photon noise sources shows that ground-based bolometers (dashed green) can achieve performance close to fundamental limits on CMB noise. The reported noise for the Planck CMB satellite are shown as squares (coherent receiver) and stars (bolometers).  Bolometer (ground) noise is based on 20\% fractional bandwidth and atmospheric noise for Chile or South Pole at a zenith angle of $45^{\circ}$. Figure from Ref. \citenum{hanany_cmb_2013}.}
	\end{figure}

	While the CMB can be observed over a wide range of frequencies, other emissions, originating from unrelated astrophysical processes, superimpose on CMB temperature and polarization anisotropies in CMB observations, in a frequency-dependent way\cite{2009LNP...665..159D}, as shown in Figure \ref{fig:cmb_spectrum}. Among those, emissions from the interstellar medium of our own Galaxy, the Milky Way, are a strong nuisance for cosmological analyses. The subtraction of this contamination requires methods that exploit multi-frequency observations of the microwave sky. Models of multi-component sky emission are used to generate mock observations and evaluate the performance of component separation methods\cite{2008A&A...491..597L,2009A&A...503..691B,2009AIPC.1141..222D}. From analyses with existing instruments that address these issues, the need is clear for CMB observations across a wide frequency domain, ranging at least from about a few tens of GHz to a few hundreds of GHz, using a combination of spaceborne and ground-based instruments. Concepts for space missions have been studied, submitted to space agencies, and in cases already partially approved by funding agencies\cite{2018JCAP...04..014D,2019arXiv190210541H,2020SPIE11443E..2FH}. In parallel, a robust ongoing program for CMB observations is under way. So-called ``stage 3'' experiments, deploying tens of thousands of cryogenically cooled bolometers at millimeter wavelengths, are being deployed in dry, high-altitude observatories to minimize noise originating from the Earth's atmosphere\cite{2019JCAP...02..056A,2020SPIE11453E..14M,2020SPIE11453E..2AS}. 
	From these high, dry sites, cryogenically cooled bolometers are able to achieve noise performance that is superior to coherent receivers and close to fundamental photon noise limits for a range of microwave frequencies, as shown in Figure \ref{fig:cmb_spectrum}.
	The next step for ground-based observations will be the deployment of hundreds of thousands of bolometers in a collection of telescopes observing at the South Pole and in Chile. This ultimate ground-based observatory for primordial CMB science is the ``stage-4'' effort for CMB observations from the ground, known as CMB-S4\cite{2019arXiv190704473A,2022arXiv220308024A}.

	\subsection{CMB-S4: The next-generation ground-based cosmic microwave background experiment}
	
	CMB-S4 is designed to significantly advance the sensitivity of CMB observations, and in doing so to enhance our understanding of the origin, matter content, and evolution of the Universe from the time of cosmic inflation to the
	growth of structure throughout cosmic history. CMB-S4 is designed following four main science drivers. The first is the quest for primordial B-modes, at a level allowing for either detecting primordial gravitational waves from inflation, or ruling out a large range of inflationary models. The second is the search for additional light particles beyond the three light neutrinos of the standard model, looking for their subtle impact on the energy density in the early universe. The third is mapping the formation of large scale structures, by detecting the emergence of clusters of galaxies across cosmic time. Finally, the fourth science driver is the search for transient emissions from stars, active galactic nuclei, and mm-wave counterparts to gamma-ray bursts.
	
	To conduct this science program, CMB-S4 will deploy telescopes at two sites, one located at the South Pole station in Antarctica, and the other in the Atacama desert in Chile. The instrument must map fluctuations of incoming power of the order of a few $10^{-19}$ Watts in arcminute-size pixels over large fractions of sky. The sensitivity of detectors observing from the ground is limited by the photon noise from the terrestrial environment, i.e. by fluctuations in the number of incoming photons in the frequency band of observation, which include not only in-band photons of astrophysical origin, but also photons from the Earth's atmospheric emission and from the warm optics of the instrument. To mitigate this photon noise, CMB-S4 will observe in wide frequency bands for increased S/N ratio (to first order, the signal increases proportionally to the bandwidth $\Delta \nu$, while the photon noise is proportional to $\sqrt{\Delta \nu}$) and deploy hundreds of thousands of independent detectors, observing repeatedly the same areas of sky for several years of operation. Small telescopes will be focusing on large-scale measurements of the primary CMB polarization B-modes, while large-dish telescopes will observe small scale primary CMB intensity and polarization fluctuations, as well as secondary signals due to the imprint of the cosmic web and of galaxy clusters on the CMB anisotropy and polarization patterns. 
	
	The need to identify and separate emission from non-cosmological sources, and in particular emission originating from the Galactic interstellar medium, demands that CMB-S4 observe in the full 20-300\,GHz frequency range, with the exception of frequency ranges affected by strong absorption and emission by oxygen and water vapor in the Earth's atmosphere. At low frequency, synchrotron emission dominates, and can be measured accurately for subtraction of its emission in CMB observations in atmospheric windows centered around 90\,GHz and 150\,GHz, where the sensitivity to CMB emission is the best. At high frequencies, thermal emission from interstellar dust strongly dominates, and can be measured accurately for CMB cleaning. The need for wide frequency bands in a broad frequency domain extending from 20 to 300\,GHz means that any emission from human-made telecommunication systems in this frequency range is a potential source of contamination for the very sensitive CMB observations with CMB-S4.

	\begin{table}[]
	\centering

    \begin{tabular}{lllll}
    \multicolumn{3}{l}{\textbf{Large-aperture telescopes (LAT)}}                          &              &                                  \\
                                                            & \textit{ULF} & \textit{LF}  & \textit{MF}  & \textit{HF}                      \\ \hline
    \multicolumn{1}{|l|}{\textbf{Band center(s) {[}GHz{]}}} & 20           & 26 / 39      & 92 / 149     & \multicolumn{1}{l|}{227 / 286}   \\
    \multicolumn{1}{|l|}{\textbf{Fractional bandwidth}}     & 0.25         & 0.33 / 0.45  & 0.32 / 0.28  & \multicolumn{1}{l|}{0.26 / 0.21} \\ \hline
                                                            &              &              &              &                                  \\
    \multicolumn{3}{l}{\textbf{Small-aperture telescopes (SAT)}}                          & \textbf{}    & \textbf{}                        \\
    \textbf{}                                               & \textit{LF}  & \textit{MF1} & \textit{MF2} & \textit{HF}                      \\ \hline
    \multicolumn{1}{|l|}{\textbf{Band center(s) {[}GHz{]}}} & 26 / 39      & 85 / 145     & 95 / 155     & \multicolumn{1}{l|}{227 / 286}   \\
    \multicolumn{1}{|l|}{\textbf{Fractional bandwidth}}     & 0.33 / 0.45  & 0.24 / 0.22  & 0.24 / 0.22  & \multicolumn{1}{l|}{0.26 / 0.21} \\ \hline
    \end{tabular}
    \vspace{5mm}

    \caption{Current design targets for band centers and fractional bandwidth for the CMB-S4 experiment. The LAT and SAT telescopes have differing measurement requirements based on the overall experiment's science goals, and do not have identical frequency bands. The band designations \textit{ULF, LF, MF, HF} represent distinct detector wafer types. All detector wafers except the ULF will have dichroic pixels, with detectors sensitive to two different frequencies bands in each pixel.}
    \label{tab:cmbs4}
    \end{table}
	
	
	Because CMB-S4 is a large project and a significant scientific investment, the design of the experiment is driven by requirements on its science goals and measurement capabilities, and the resulting specifications on its technical performance.
	CMB-S4 will use the demonstrated technologies and techniques summarized in Section \ref{sec:instrumentdesign} to dramatically increase its sensitivity while maintaining strict control of instrumental systematic effects.
	Forecasting for CMB-S4’s constraints on primordial gravitational waves is extremely complex, and the best estimates described in the CMB-S4 collaboration’s recent forecasting paper\cite{cmbs4forecasting_2022} are based on scaling of past performance of CMB instruments, which attempt to capture subtle systematic effects and degradations in sensitivity in this difficult measurement.
	The details of a reference instrument design capable of achieving CMB-S4's science goals are given in Ref. \citenum{cmbs4_dsr}.
	This design continues to be refined by the CMB-S4 collaboration and project team. 
	The most relevant parameters for this work are the frequency bands for the small- and large-aperture telescopes for CMB-S4, which are given in Table \ref{tab:cmbs4}.
    Updates on recent design developments for CMB-S4 can be found in concurrent proceedings of this conference, including an overview of the current design in Ref. \citenum{spie_overview_2022}, the design of the detectors and readout system in Ref. \citenum{spie_drm_2022}, optical design of the large aperture telescopes in Ref. \citenum{spie_lat_2022}, and sidelobe modeling and mitigation in Ref. \citenum{spie_sidelobes_2022}.
    

	\subsection{Existing Protections and Interference Criteria for Radio Astronomy} 
	
	Internationally, use of the radio spectrum for commercial, public, and scientific applications is governed by the Radio Regulations (RR), an international treaty followed by most national administrations.  
	An agency of the UN, the International Telecommunication Union (ITU), in particular its Radiocommunications sector (ITU-R)\footnote{\url{https://www.itu.int/en/ITU-R/information/Pages/default.aspx}}, is responsible for the ongoing evolution of the RR in response to changes in technology and the competing needs of spectrum users.  
	National administrations maintain consistency with the RR where required under treaty; this applies especially to activity that has effects beyond national borders.  Licensing authority over domestic and foreign active spectrum users is exercised independently by administrations within national boundaries.  For example, the FCC has granted “authorizations” to US-based satellite networks SpaceX Starlink and Amazon Kuiper, and a “market access grant” to UK-based OneWeb.  For these large constellations in low Earth orbit (LEO), uniform technical and operational characteristics worldwide are desirable, but individual administrations can in principle impose different licensing conditions.
	
	Within the RR, Radio Astronomy (known as the Radio Astronomy Service, or RAS) is granted protection from harmful interference within certain bands covering a small percentage of the radio spectrum.  The level of interference that is harmful to a RAS observation depends on the observing mode---definitions of harmful interference for single-dish, connected interferometer, and VLBI observations are given in ITU-R recommendation RA.769.  These bands allocated to radio astronomy are often associated with the rest frequencies of important atomic and molecular spectral lines, with bandwidth sufficient to cover the modest Doppler shift associated with sources in the local galactic neighborhood.  However, while the RAS bands continue to be of great value, much of modern radio astronomy and experimental cosmology has outstripped this protection, either because it observes spectral line emission at cosmological redshifts, or because very wide continuum bandwidths are needed to achieve required sensitivity. 
	Many modern instruments including CMB-S4 must necessarily use unprotected spectrum allocated to active services.
	
	
	
	

	\section{CMB Instrument Overview}\label{sec:instrumentdesign}
	
	
	Here we summarize  
	aspects of CMB instrument design most relevant to 
	radio frequency interference susceptibility, with a focus on their planned implementation in CMB-S4.
	A detailed and comprehensive description of the state-of-the-art technologies for ground-based CMB instruments 
	can be found in the CMB-S4 Technology Book\cite{cmbs4_technologybook}, the CMB-S4 Decadal Survey Report\cite{cmbs4_dsr}, as well as in the references throughout this section.
	Relevant details on how CMB instruments conduct observations and how the data is analyzed are given in Section \ref{sec:analysis_mitigation}.

	\subsection{Background-limited sensitivity with transition edge sensor bolometers}\label{ssec:bolometers}
	
	Bolometers are detectors that measure incident power through changes in temperature.
	Each bolometer consists of an absorber connected to a thermal reservoir through a weak thermal link, and an attached sensor that has a temperature-dependent electrical resistance. 
	Any incident radiation, whether electromagnetic or particle radiation, raises the temperature of the device.
	For millimeter to sub-millimeter wavelengths (electromagnetic radiation between approximately 100 GHz and 1 THz), bolometers are exceptionally sensitive and able to achieve performance limited primarily by photon shot noise\cite{richards1994}. 
	In this regime, 
	the bolometer is operated at sub-Kelvin temperature and the noise from the detector itself is subdominant to 
	photon noise from incident light.  
	
	
	Transition-edge sensor (TES) bolometers take advantage of the behavior of a superconducting film in the transition between normal and superconducting, where small temperature fluctuations result in significant changes in electrical resistance. 
	The linearity, dynamic range, and time constant of a transition-edge sensor bolometer can be greatly extended through feedback. 
	When a constant voltage bias is applied to the TES,  the deposited electrical power is inversely proportional to resistance.  
	This results in strong negative electrothermal feedback, where electrical power decreases with incident radiative power, keeping total power on the TES constant.  Readout of the TES bolometers is accomplished by coupling the detectors to Superconducting Quantum Interference Devices (SQUIDs), which can be used as exceptionally sensitive magnetometers.
	The current from the TES is inductively coupled to the SQUID’s input coil and induces a small magnetic field, which induces a voltage in the SQUID. 
	The technique of voltage-biased superconducting transition edge sensors read out with low-impedance SQUID readout has become a powerful technique not just for precision cosmic microwave background measurements, but is a widely used technique in millimeter-wave astronomy as well as photon-counting applications at shorter wavelengths\cite{Irwin1995,IrwinHilton2005} and dark-matter detection \cite{CDMS_1999}.
	
	In addition to the excellent sensitivity of this technique, these superconducting electronics are also readily scalable using advanced silicon fabrication processes\cite{cmbs4_technologybook}. 
	The instantaneous sensitivity of instruments using background-limited detectors is increased only through increasing detector counts 
	and significant technology developments have led to several successive order of magnitude leaps in detector counts and sensitivity over the past two decades.
	This has resulted in the successful deployment of large cryogenic focal planes with thousands and tens of thousands of bolometers on several recent instruments\cite{sobrin_spt3g_2022, westbrook_2018, moncelsi_receiver_2020, henderson_advact_2016}, with large fields-of-view and high instantaneous sensitivity. 
	
	
	Large cryogenic TES arrays have driven the development of sophisticated multiplexing techniques\cite{henderson_advact_2016,bender2020,mccarrick2021} that read out dozens of bolometers on a single pair of wires. 
	The success of these techniques enables the design of CMB-S4, which will further scale detector counts by another order of magnitude, to $\sim$500,000 TES bolometers deployed on several telescopes, occupying several square meters of cryogenic focal planes.  
	For CMB-S4, these detectors will be fabricated on 150-mm silicon wafers, with the wafer types listed in Table \ref{tab:cmbs4}.
	The detector wafers are integrated into a modular package with associated readout electronics.
	Each wafer module will undergo significant characterization and screening to ensure that it meets all performance requirements.
	The production of such a large number of precision superconducting devices represents a significant effort that will take approximately 3 years across multiple fabrication sites\cite{spie_drm_2022}.


	\subsection{Optical Coupling to Focal Plane and Photometric Passband Design}\label{ssec:opticalcoupling}

\begin{figure} [ht]
		\begin{center}
			\begin{tabular}{cc} 
			    \includegraphics[height=4.1cm]{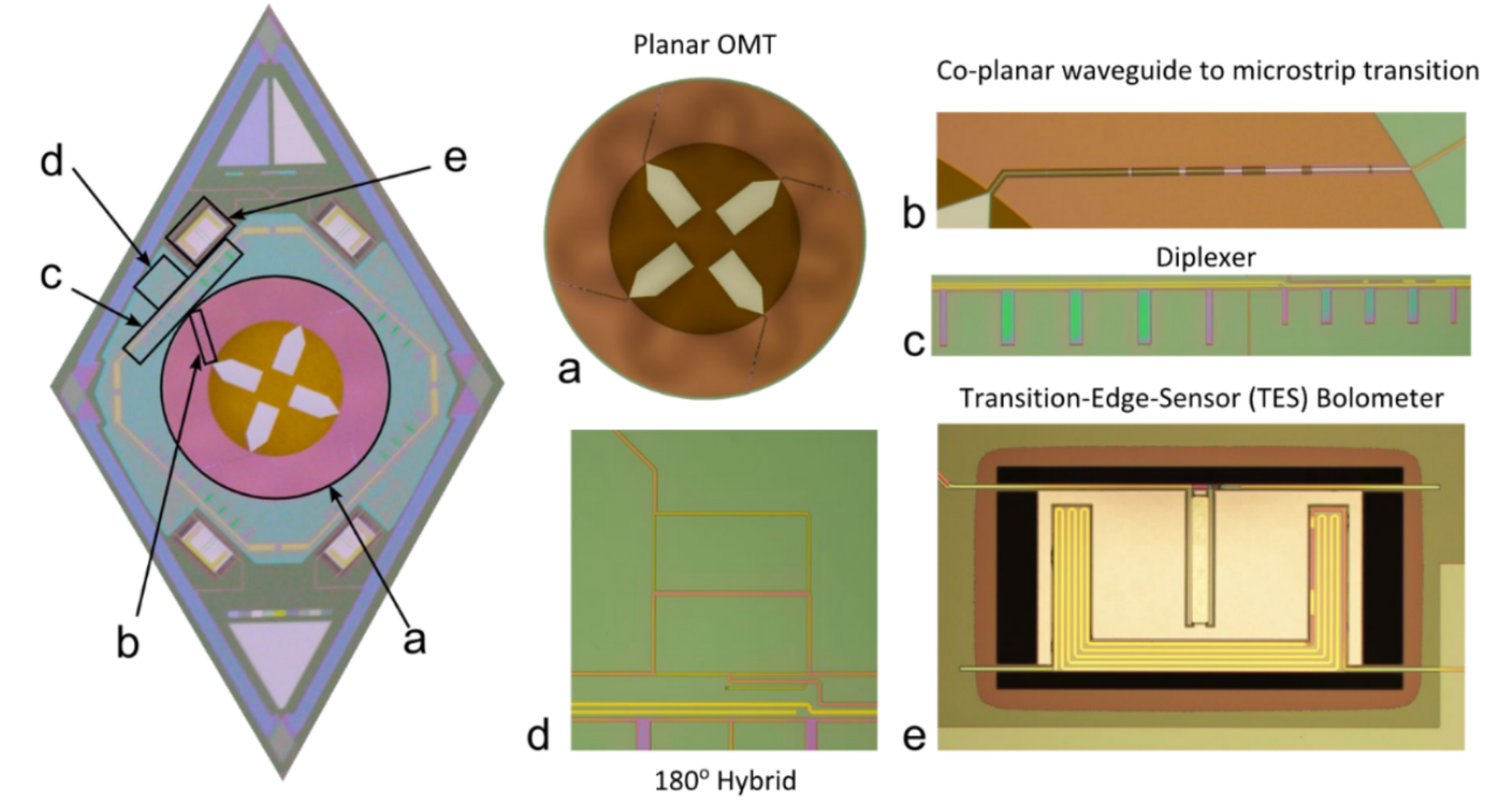}
				\includegraphics[height=4.5cm]{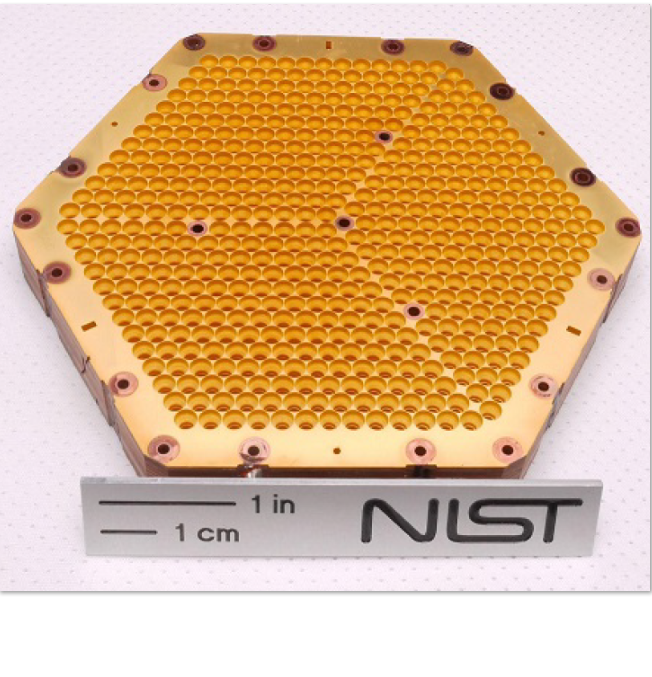}
				\includegraphics[height=4.8cm]{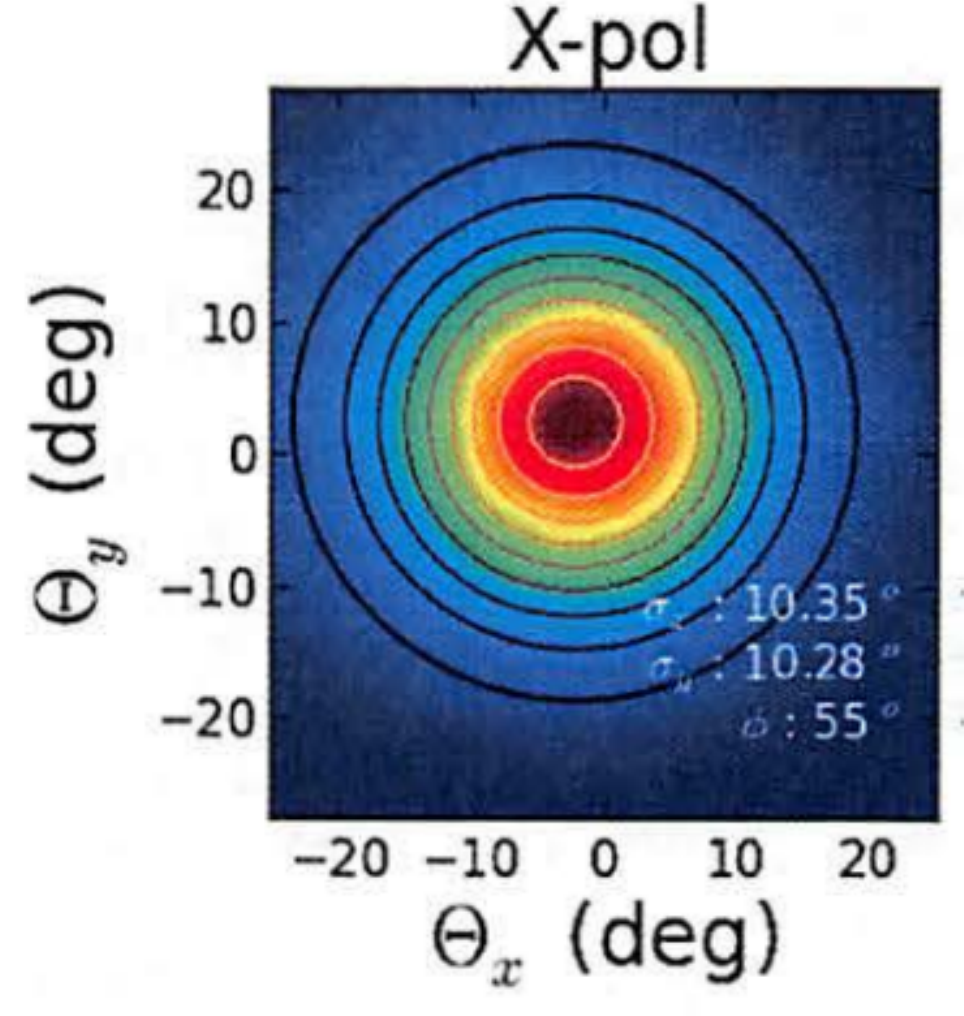}
			\end{tabular}
		\end{center}
		\caption[actbeams] 
		{ \label{fig:ACT_beams} 
			(Left) Image of a single pixel from the Advanced ACTPol CMB instrument with major features labeled and magnified\cite{cmbs4_dsr}. (Center) A photograph of a CMB feedhorn array, with 503 spline-profiled feeds\cite{hubmayr_feedhorn_2015}. (Right) The 2D angular response measurement of a horn-coupled 150 GHz detector, with X polarization beams plotted on a linear color scale along with fits to a 2D Gaussian profile\cite{hubmayr_feedhorn_2015}.  
			}
	\end{figure}

	TES bolometers are exceptionally sensitive to incident power, and the optical coupling of these detectors to the sky is carefully designed to maximize sensitivity and instrument performance. 
	Each pixel in the array has a microwave element that couples to incident radiation. 
	Feedhorn coupling is a widely-used technique in CMB instruments\cite{feedhorns1985,hubmayr_feedhorn_2015,simon2016,rostem2016,cmbs4_dsr} 
	A horn antenna defines the angular response on the sky and a planar ortho-mode transducer (OMT) probe couples the incident microwave power to the detector.
	This configuration, shown in Figure \ref{fig:ACT_beams}, creates a well-defined angular response that is quite symmetric between polarization orientations. 
	
	
\begin{figure} [ht]
		\begin{center}
			\begin{tabular}{cc} 
				\includegraphics[height=4cm]{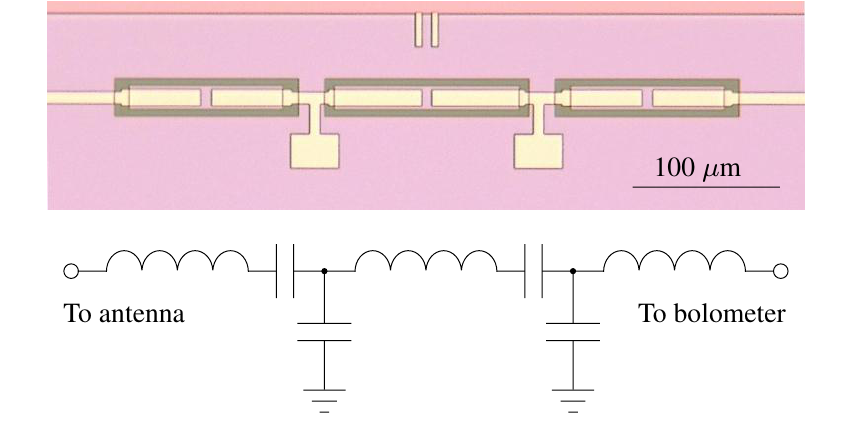}
				\includegraphics[height=4.5cm]{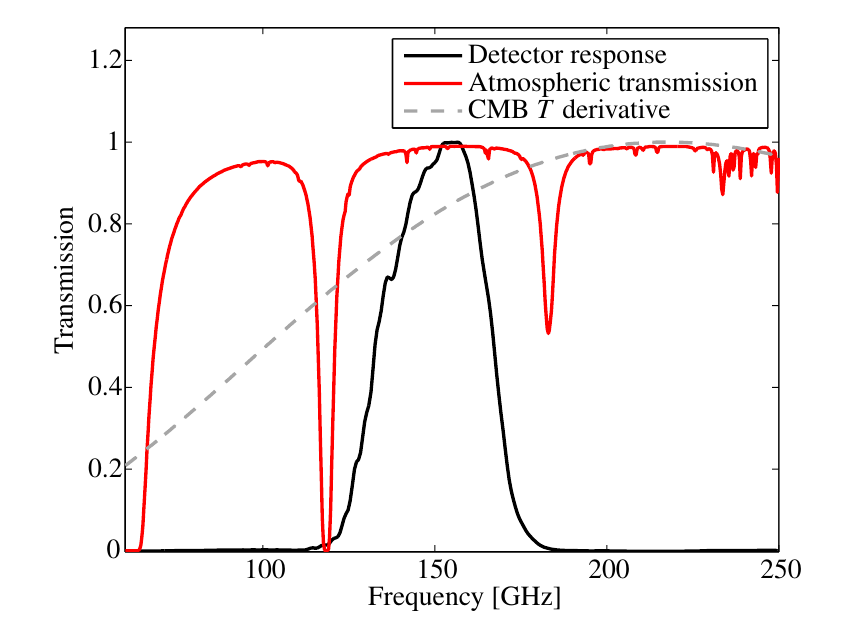}
			\end{tabular}
		\end{center}
		\caption[actbeams] 
		{ \label{fig:bk_bands} 
			(Left) The on-chip band-defining filter and equivalent circuit for BICEP2 defines a frequency band centered at 150 GHz with a 25\% fractional bandwidth.  (Right) The resulting measured frequency response spectrum (black solid line) is overlaid with typical atmospheric transmission at the South Pole (red). Both figures from Ref. \citenum{bicep2_2014_ii}.}
	\end{figure}

	Similar to astronomical measurements in the optical, ground-based measurements of the CMB use a relatively standard set of photometric passbands. 
	These passbands are constrained 
	by atmospheric transmission windows defined by absorption lines of molecular oxygen and water vapor.
	An example of this is shown in Figure \ref{fig:bk_bands}. 
	Instruments are designed to maximize sensitivity through broad bandwidth while avoiding these absorption lines.
	The photometric passbands are defined by planar microwave structures 
	between the antenna and the TES bolometer. Figure \ref{fig:bk_bands} shows an example RF filter and the resulting passband for the BICEP2 instrument\cite{bicep2_2014_ii}.
	Each TES detector has its own individual on-chip RF filter defining the frequency response and slight variations between theoretically identical filters are introduced in the fabrication process.  Additionally, filters cannot be modified after fabrication.
	Power measured by the TES after the filter and further read out contains no frequency information. 
	Passbands are characterized through 
	Fourier-transform spectroscopy, measuring the detailed frequency response of each detector for use in data analysis (for examples see Refs. \citenum{matsuda_fts, pan2018}). 
	
	Precision measurements of the cosmic microwave background must utilize all available atmospheric transmission windows around the CMB spectral peak at 160 GHz to characterize and remove foreground signals, such as those from the Galaxy, which have different spectral signatures. 
	Multi-chroic pixels are a means to further increase sensitivity and frequency coverage for a fixed focal plane size, with each single pixel having multiple detectors with different passbands and polarization sensitivity. One example of a large-format multi-chroic TES array is the SPT-3G experiment, which is currently installed on the South Pole Telescope.\cite{sobrin_spt3g_2022}  
	The broadband optics design for SPT-3G results in a 430 mm diameter image plane, which is filled with 2690 pixels.
	Each pixel separates the incoming microwave light by orthogonal linear polarization and then again into three different passbands centered at 95/150/220 GHz. 
    These pixels triple the sensitivity for the same number of optical coupling elements, while also providing crucial frequency information for disentangling the aforementioned foreground signals.
    With this sensitive, broadband design, the SPT-3G cryogenic focal plane contains over 16,000 detectors and associated readout, and weighs 22 kg.
	
	CMB-S4 will similarly utilize multi-chroic pixels to maximize the sensitivity of each focal plane.
	Most wafer types (listed in Table \ref{tab:cmbs4}) will include di-chroic pixels.
	Wafers of different types and frequency bands can be placed in the same focal plane to maximize frequency coverage of a single instrument.
	
	



	\subsection{Characteristics of Established CMB Observing Sites}\label{ssec:sites} 
	
	As noted above, CMB observations take place in the millimeter-wave atmospheric transmission windows between strongly absorbing water vapor and molecular oxygen emission lines.  Within these windows, the atmospheric opacity is dominated by the line wing and continuum absorption of water vapor, with lesser contributions from dry air continua and trace species such as ozone.  The water vapor overburden above the Earth's surface ranges over roughly two orders of magnitude depending on local climate and surface altitude, from hundreds of microns to tens of millimeters of precipitable water vapor.  This is a huge factor considering that this column density affects absorption exponentially; this is why sites that combine dry climate with high altitude are critically valuable for ground-based radio astronomy.
	
	Contemporary advances in ground-based experimental cosmology have been made possible by taking advantage of the unusual meteorological characteristics and support infrastructure of two dry, high-altitude sites: the Amundsen-Scott South Pole Station and locations within the Chajnantor Science Reserve in the high Atacama Desert in northern Chile including Cerro Toco.  Both are in desert zones associated with descending dry circulation in the global water cycle, augmented at the Pole by the extreme cold and in the Atacama by the Andes mountains, which serve as a barrier to moist circulation and afford access to especially high altitudes.  In other characteristics, these two sites are highly complementary: the South Pole atmosphere is extremely stable, minimizing atmospheric fluctuation noise in CMB observations, whereas the Atacama offers access to a larger fraction of the sky at the cost of greatly reduced atmospheric stability owing to the strong diurnal forcing.  CMB-S4, as envisaged, will deploy instruments to each of these sites to exploit the advantages of both.
	


	\section{RFI Coupling Mechanisms and Instrument-Level Mitigations}\label{sec:coupling}
	




	Deploying large arrays of broad bandwidth, background limited detectors yields instruments with high instantaneous sensitivity. 
	However, characterizing faint cosmological signals requires not just raw sensitivity, but also strong control and understanding of instrumental and other systematic effects on the measurement. 
	For this reason, the instrument design includes many features for the mitigation of unwanted signals and stray radiation that could contaminate the measurement.
	Sources of stray radiation are not limited to RFI, but also include thermal emission from the instrument and surrounding terrain, emission from astronomical sources including the Sun, Moon, and Galaxy, and particle radiation from cosmic rays.
	In this section we describe the coupling mechanisms for stray radiation with a focus on coupling of RFI, and discuss mitigation techniques that prevent or 
	reduce the effective coupling to the instrument.

	It is useful to distinguish between different modes of RFI coupling into the telescope.  
	A first distinction is between RFI that is absorbed by the detectors themselves (direct coupling) versus RFI coupling to the supporting wiring and readout electronics (indirect coupling).  
	RFI coupled directly to the detectors can be further divided into in-band and out-of-band absorption relative to the intended observing band of a given detector.
	Here we also further distinguish between in-band direct coupling for RFI within the telescope's astronomical signal path (direct coupling to main beam) and RFI originating outside of the telescope's field of view (direct coupling to stray light).

	We discuss these roughly in order of increasing coupling strength, starting with indirect coupling, followed by direct coupling to out-of-band radiation, direct coupling to stray light, and finally direct in-band coupling to the main beam.
	We describe the special case of coupling to satellite transmitters in Section \ref{sec:satelliterfi}.
	For direct in-band coupling to the main beam (which is the astronomical signal path), any RFI such as fundamental or harmonic emissions from a transmitter will reach the detector unimpeded. Mitigation techniques for this type of RFI are found in Section \ref{sec:analysis_mitigation}.

	\subsection{Indirect Coupling}

	Indirect coupling refers to RFI that is not absorbed by the detectors themselves.
	This can occur through a variety of mechanisms throughout the instrument, and in some instances can be difficult to distinguish from direct coupling to out-of-band radiation (discussed in \ref{ssec:outofband}).
	The low noise readout electronics for the detectors can be especially sensitive to electromagnetic interference, including the SQUID amplifiers and room temperature electronics.
	Indirect coupling is an important consideration for overall instrument design, since sensitive electronics and their cabling are located alongside many other auxiliary electronics and power systems for the instrument and telescope.


	Although there are many complex mechanisms for indirect coupling of RFI, many external coupling paths to the backend electronics can be mitigated through standard shielding techniques, and indeed many of the measures taken to minimize coupling to locally-generated RFI are effective against external RFI sources.
	Indirect coupling through electromagnetic radiation incident on the optical path is more difficult to shield against, and is discussed in \ref{ssec:outofband}.
	Portions of the cryogenic receiver that are behind the detectors and optical path can be effectively enclosed in an RF quiet volume with the metal vacuum shell forming the bulk of the shielding. 
	For all wiring entering and leaving the receiver's otherwise RF-tight volume, in-line filters are used along with shielded cables.
	Typical shielding for a CMB instrument includes not just RF shielding but also significant cryogenic magnetic shielding to suppress the Earth's magnetic field at the SQUID amplifiers.
	CMB-S4 will benefit from significant experience in effective electromagnetic shielding of sensitive receiver electronics.
	Requirements on susceptibility to indirect coupling and electromagnetic compatibility of sub-systems will inform the shielding and grounding designs for the receiver, telescope, and associated electronics for CMB-S4.


	\subsection{Direct Coupling to Out-of-Band Radiation}\label{ssec:outofband}

	Direct absorption of out-of-band radiation incident on the bolometers can, in principle, occur at any frequency, as described in Section \ref{ssec:bolometers}.
	In practice, only out-of-band radiation at lower frequencies is a significant concern.
	Out-of-band radiation at higher frequencies, including far-infrared and optical radiation, can be thoroughly suppressed by the short-wave blocking filters engineered into the cryostat\cite{halpern86,ade2006,ahmed2014}, that do not substantially impact in-band radiation.
	Spectrally-adjacent out-of-band radiation is
	suppressed by the same band definition filters described in \ref{ssec:opticalcoupling} that must robustly reject adjacent atmospheric emission line radiation.  
	
	A remaining RFI hazard that has been significant for current-generation CMB experiments is out-of-band radiation at frequencies several times lower than the observing frequency but above the waveguide cutoff frequency of the telescope aperture, which is about 350 MHz for a 0.5 meter aperture diameter.  At these frequencies, the absorbing baffling within the telescope tube is not very effective, and radiation can propagate as guided or conducted modes within the cryostat structure.  The associated sensitivity pattern is roughly isotropic outside the telescope aperture and thus susceptible to ground-level transmitters.
	Indeed, as discussed below, interference to BICEP-3 from UHF communications radios at Amundsen-Scott station severely degraded BICEP-3 data until it was mitigated through technical and operational measures.


	The horn-coupled detector design that will be be used for CMB-S4 will offer inherent shielding against below-band radiation on the front side of the focal plane that is below the waveguide cutoff frequency at the throat of the feedhorns.  In addition, the CMB-S4 detectors and sensitive readout electronics will be packaged in enclosed modules behind the feedhorn plates, which will provide back side shielding against stray RFI within the cryostat.  Wideband testing of achieved RFI immunity will be important to ensure that weakly-coupled leakage through electrical feedthroughs and other required breaks is not resonantly enhanced within these enclosed modules.

	\subsection{Direct Coupling to Stray Light}\label{ssec:sidelobes}

\begin{figure} [ht]
		\begin{center}
			\begin{tabular}{cc} 
				\includegraphics[height=5cm]{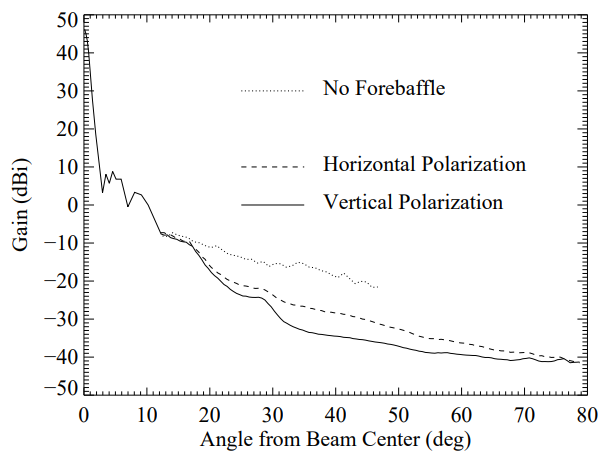}
				\includegraphics[height=5cm]{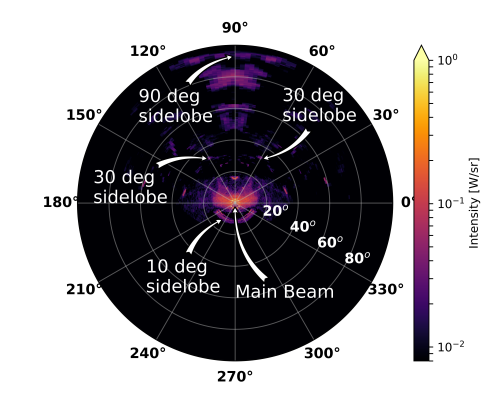}
			\end{tabular}
		\end{center}
		\caption[sidelobes] 
		{ \label{fig:sidelobes} 
			(Left) The azimuthally-averaged measured sidelobe response for the BICEP small aperture CMB instrument is shown, demonstrating additional attenuation achieved with the use of an absorptive forebaffle. Figure from Ref. \citenum{takahashi_characterization_2010}. (Right) A map of simulated sidelobes from raytracing of the large aperture Atacama Cosmology Telescope is overlaid with arrows showing the position of measured far sidelobes for the instrument, at the level of approximately -20 to - 40 dBi. Figure from Ref. \citenum{gallardo_far_2018}.}
	\end{figure}

	CMB instruments are designed to be as well shielded from stray radiation as possible, while maintaining a relatively large field of view on the sky (10--30 degrees).  
	Minimizing telescope sidelobes is an important design consideration, since stray radiation from the ground, Sun, Moon, and our Galaxy causes systematic effects that can contaminate the measurement. 
	The telescopes generally have large co-moving shields, and many unwanted sidelobes can be terminated on an absorbing surface or reflected onto the cold sky. 
	For small aperture CMB telescopes, this shielding can be very effective to eliminate coupling far from the main beam, reaching shielding factors of -40 dBi more than 40 degrees away, as shown in Figure \ref{fig:sidelobes}.  
	For larger aperture CMB telescopes (D $\sim$ 6 meters), designed to survey the entire sky, designing effective co-moving shielding that does not interfere with the telescope structure and can withstand harsh environmental conditions is significantly more difficult. 
	
	Because of this difficulty in effective shielding, current large aperture CMB telescopes have measured strong sidelobes far from the main beam, arising from scattering and reflections from optical surfaces. 
	Modeling sidelobes in these complex optical systems is difficult and not always accurate, so prior experience heavily informs expectations in future instruments. 
	Sidelobe features will be characterized in CMB-S4 to determine their location and strength, but may not be well known at the start of observations.  
	A bright object passing through one of these far sidelobes may couple strongly and be difficult to identify.

	Within the telescope’s field of view, tens of degrees from the main beam, stronger near-sidelobe features can arise from optical crosstalk within the camera, or electrical crosstalk between detectors. Electrical crosstalk results from indirect coupling between detector wiring or readout, but appears in the measurement as additional in-band signal. 
	These features are sometimes unavoidable in the complex instrument, but can be carefully characterized and corrected for in analysis.  
	Objects passing through these near sidelobes would also be within the main beam of nearby detectors, providing multiple coincident observations to track and understand the source of emission.


	\subsection{Direct In-Band Coupling to Main Beam}\label{ssec:direct_mainbeam}
	
	The most straightforward coupling case is in-band RFI absorbed by the detectors.  
	The directional sensitivity to in-band RFI will be dominated by coupling via the main beam
	antenna pattern associated with the victim detector.  
	As this is the astronomical signal path, any RFI such as fundamental or harmonic emissions from a satellite transmitter will reach the detector unimpeded.
	For known sources of emission, including the Sun and Moon, telescope scan strategies are designed to avoid coupling of these sources to the main beam or near sidelobes during science observations.
	For example, observations can be planned around Sun avoidance of at least 30 degrees.
	The Moon can be avoided similarly, or its position can be included in detailed analysis tests for contamination (further discussed in Section \ref{ssec:efficiency})
	However, avoidance strategies that are effective for sparse sources that are nearly stationary in the sky are infeasible for numerous in-motion sources such as those associated with a large LEO constellation.
	For such sources, direct coupling of in-band emission to the main beam
	will be unavoidable.
		
	\section{Historical Examples: RFI Mitigation Efforts at South Pole Dark Sector}\label{sec:rfi_pole}
		

    Historically, locally-generated harmful interference has been experienced by the BICEP/Keck series of CMB experiments at Amundsen-Scott South Pole Station.  Here we briefly summarize two instructive examples of RFI sources and successful mitigation measures. Both examples also demonstrate the sensitivity of CMB instruments to out-of-band radiation below the instrument's measurement band.
    
    The first example was interference from the 450 MHz UHF land mobile radio (LMR) system used to support operations and safety at the Pole.  When BICEP3 was deployed in 2015, two factors conspired to produce debilitating RFI for the new telescope.  First, BICEP3 had a significantly larger cryostat diameter (0.7 m) than earlier BICEP/Keck instruments (0.25 m), lowering the cutoff frequency from 500 MHz to 260 MHz.  Second, the transmission power of the LMR base station had been increased to accommodate increased building entry loss associated with a newly-installed metal roof on the station main building.  The intermittent nature of the transmissions resulted in months of data loss before the RFI source could be traced.  A solution was adopted in coordination with NSF and US Antarctic Program spectrum managers: the omnidirectional base station antenna on the main building roof was replaced with one having -22.5 dBi gain towards the "Dark Sector" where the astronomical facilities are located, and additionally transmit power is reduced when outdoor maintenance operations are not occurring there.
    
\begin{figure} [ht]
		\begin{center}
			\begin{tabular}{cc} 
				\includegraphics[height=5cm]{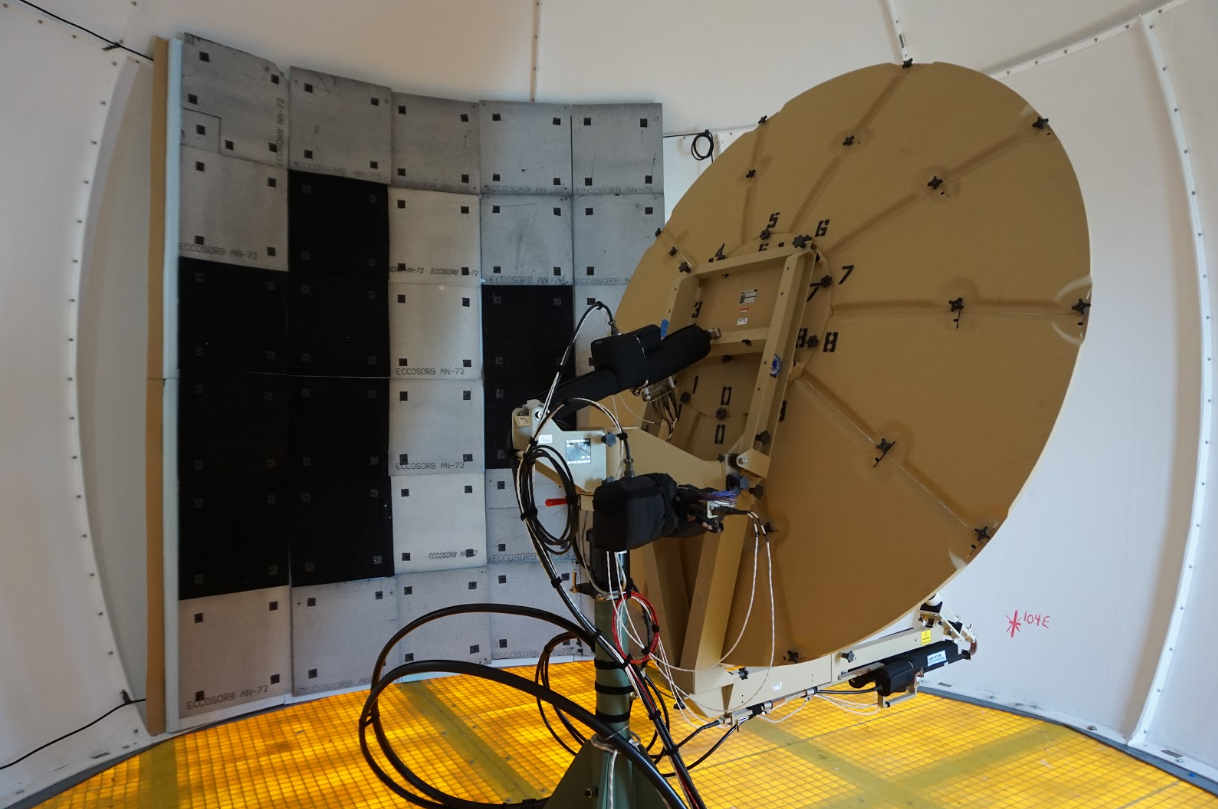}
				\includegraphics[height=5cm]{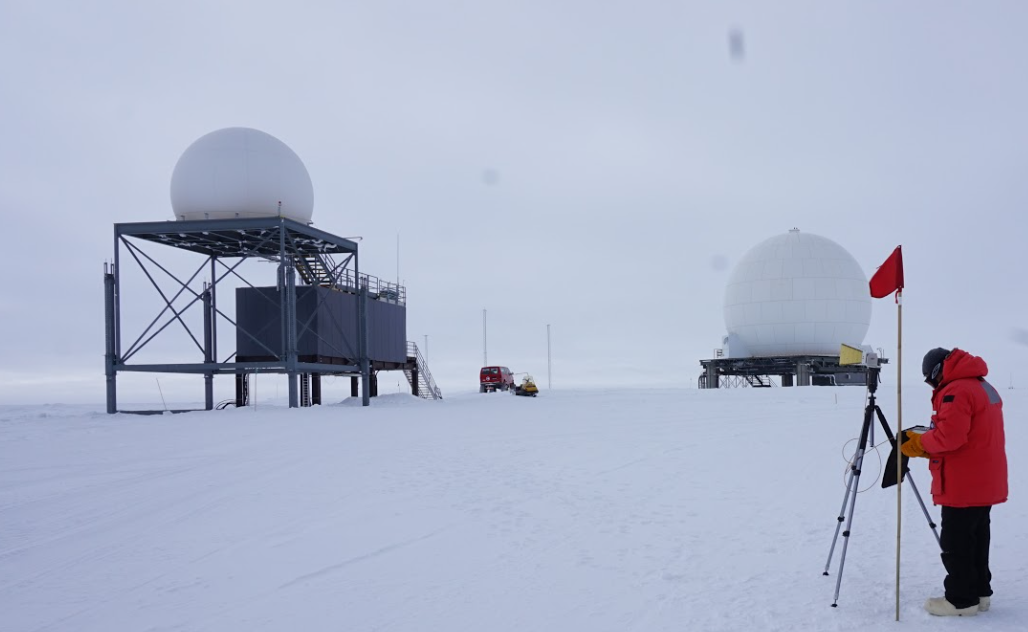}
			\end{tabular}
		\end{center}
		\caption[rfimitigation] 
		{ \label{fig:pole_mitigation} 
			 In coordination with NSF and US Antarctic Program spectrum managers, transmitters at South Pole are carefully controlled to limit RFI in the Dark Sector where CMB telescopes are sited. (Left) A satellite uplink transmitter is shielded in the direction of the Dark Sector using panels of microwave absorber. (Right) Spectral survey measurements, shown here using a standard gain horn and portable spectrum analyzer, are used to confirm levels of RFI from communication facilities in directions toward the Dark Sector are below levels determined to contaminate telescope data. }
	\end{figure}

    The second example involves measures taken to avoid harmful interference from satellite uplink transmitters at frequencies ranging from 2--15 GHz, used for data communications at the Pole.  These satellite uplink transmitters are located 1700 meters away from the Dark Sector, and must point low to the horizon towards target satellites.  Prior to mitigation, sidelobe radiation from the satcom antennas appeared as azimuth-dependent interference in the CMB telescope data streams.  Quantitative assessment of this interference led to establishment of a harmful interference threshold, which is now a working requirement for the maximum acceptable power level in this band within the Dark Sector.
    This harmful interference threshold is 10 nW/m\textsuperscript{2} integrated across 2--15 GHz, from all sources.
    This frequency range is below the measurement band of any of the instruments that were affected in the Dark Sector. Transmissions at higher frequencies could overlap with CMB measurement bands which start around 20 GHz, and would have significantly lower harmful interference thresholds.
    Reduction of power below the interference threshold has been achieved by erecting barriers faced with microwave absorber within the satcom antenna radomes, and measurements have confirmed that the current levels meet this requirement, as shown in Figure \ref{fig:pole_mitigation}.

    Broadband RFI monitor receivers are now operated continuously at each of the two main CMB observatory buildings in the Dark Sector, MAPO and DSL.  Using a combination of low- and high-frequency omnidirectional antennas located near the CMB telescopes at observatory, these monitor receivers currently record the local spectrum from 50 MHz to 12.4 GHz every two minutes, posting this data to public website.  These RFI monitor receivers have proven invaluable in identifying changing sources of RFI in the local environment, facilitating the correlation of these changes with contamination in telescope data and guiding efforts to mitigate new sources of interference.
    
	\section{Coupling to Satellite RFI}\label{sec:satelliterfi}
	
	Commercial satellite constellations for communications can comprise thousands of satellites.
	The impact of satellite constellations on astronomy has been discussed in many recent publications \cite{Tyson_2020,Hainaut2020,DQS1-2020,SATCON1-2020,Hecht2021,SATCON2-2021,DQSII-2021,Lawler2022,Mroz2022, Williams2021}, with most focused on the impact to optical astronomy.
	Optical and infrared astronomy face a significant, complex challenge because the interference from satellites results from reflection of sunlight and emission of thermal radiation, which are difficult to predict and control.
	Efforts are underway to analyze optical images affected by satellite constellations to further improve modeling efforts, aid in the development of masking techniques, and assess the accuracy of positional information currently provided by operators.\cite{SATCON2-ObsWG2021,Rawls2022}
	Growing populations of communications satellites present a different challenge to radio and millimeter-wave astronomy, where the interference results from active transmission in licensed frequency bands.
	Even existing satellite constellations pose a challenge for increasingly sensitive instruments. 
	For example, future total-power single-dish HI intensity mapping surveys with the Square Kilometer Array could be significantly impacted by transmission bands of global navigation satellite services and out-of-band leakage~\cite{Harper2018}.
	Certain current CMB instruments using coherent receivers to measure the CMB and synchrotron at lower frequencies (10--40\,GHz) are already experiencing significant interference from satellites.
	For example, the QUIJOTE CMB Experiment at Teide Observatory is impacted by contamination from downlink communications by geostationary satellites, and the new MFI2 instrument is susceptible to Starlink's Ku-Ka downlink transmissions (described in concurrent proceedings from this conference, Ref.~\citenum{quijote2022}).
	CMB instruments using the techniques described in Section~\ref{sec:instrumentdesign}, including CMB-S4, have a unique set of considerations distinguishing them from many radio astronomy counterparts, because of the broadband bolometer detectors used in these instruments.
	Although the instruments and mechanism of interference are quite different, CMB-S4 faces similar concerns to the Rubin Observatory LSST regarding insidious contamination of data for large-scale precision cosmological surveys.

	
	With the coupling mechanisms in Section \ref{sec:coupling}, many of the mitigation techniques are most effective for local sources of emission at low elevation.
	Even the historical example of harmful interference to CMB instruments from out-of-band satellite communications described in Section \ref{sec:rfi_pole} was resulting from local transmissions originating at an uplink terminal located nearby, pointed low on the horizon.
	In this section we discuss the special case of RFI from emissions from orbiting artificial satellites.
	For these celestial sources, their coupling through the astronomical signal path is very difficult or impossible to further suppress.
	The primary concerns are direct coupling of in-band radiation to a CMB detector's main beam and sidelobes.
	The susceptibility of the CMB instrument to satellite transmissions is relatively well defined by its instrument model, although some features require detailed measurements, including the far sidelobes (described in \ref{ssec:sidelobes}).
	The level of interference can be assessed as a signal-to-noise ratio using the power received by the  bolometer, and its sensitivity to incident power.
	Strong levels of interference can also exceed the bolometer's limited dynamic range (1--50\,pW, depending on the observing band), causing not just signal contamination but disruption of detector operation.
	
	Understanding the susceptibility of CMB instruments like CMB-S4 is only one piece of the complex modeling of satellite-originated RFI and its impact on survey data.
	In assessing the impact at the survey level, the strength of coupling and level of interference are just as important as the rate at which interference events occur and their duration.
	This is why emerging plans for satellite constellations are of such concern for cosmological surveys including CMB-S4.
	Assessing this emerging threat now, before the full deployment and activation of constellations, requires assumptions about active satellite transmissions like bandwidth and directivity and assumptions about orbits including altitude and inclination.
	More accurate assessment of potential interference will eventually require detailed measurements of satellite transmitters including their emissions in far sidelobes and unintentional harmonic emission.
	
	A starting point for assessing RFI from commmunications satellite constellations are the current FCC regulations on fixed satellite services. Regulatory limits for four bands are given in Table \ref{table:pfdlims}.
	For out-of-band coupling, the power flux density limits for fixed satellite services in the X band can be directly compared to the established threshold for harmful interference in the South Pole Dark Sector described in Section \ref{sec:rfi_pole}.
	The regulatory limits can also be used in combination with the CMB-S4 instrument model to assess the potential for coupling to in-band emission where the operator bands and CMB instrument bands overlap.

    \begin{table}[ht]
    \centering
    \begin{tabular}{|c|c|c|c|c|}
    \hline
    Nominal & $f_{\text{center}}$ \si{[\GHz]} & $f_{\text{low}}$ \si{[\GHz]} & $f_{\text{high}}$ \si{[\GHz]} & PFD [\si{\dB\watt\per\meter\squared}] \\ \hline\hline
    X   & 11.20 & 10.7 & 11.7 & $-116$ \\ \hline
    K   & 24.60 & 24.45 & 24.75 & $-105$ \\ \hline
    Ka  & 38.75 & 37.5 & 40.0 & $-117$ \\ \hline
    V/W & 73.50 & 71.0 & 76.0 & $-105$ \\ \hline
    \end{tabular}
    \caption{Power flux density (PFD) limits for fixed satellite services in operating bands considered within this analysis. The PFDs here are given for a 1\,\si{\MHz} reference band, and represent the maximum signal per square meter permissible from a terrestrial observer. \cite{pfdlims2010,slinkfiling2020} K, Ka, and V/W bands fall within or overlap CMB measurement bands, and X band is within the frequency range where CMB instruments have experienced harmful out-of-band interference (described in Section \ref{sec:rfi_pole}).} 
    \label{table:pfdlims} 
    \end{table}


    
    \begin{table}[ht]
    \centering
    \begin{tabular}{|c|c|c|c|c|}
    \hline
    CMB-S4 Band & Interfering Band & Bandwidth [dB MHz] & Telescope Forward Gain [dBi] & Power [pW]  \\ \hline\hline
    SAT 26\,GHz & K   & 25 & 40 & $10^4$   \\ \hline
    LAT 26\,GHz & K   & 25 & 60 & $10^6$  \\ \hline
    SAT 39\,GHz & Ka  & 34 & 40 & $10^3$   \\ \hline
    LAT 39\,GHz & Ka  & 34 & 60 & $10^5$   \\ \hline
    SAT 85\,GHz & V/W & 31 & 50 & $10^5$   \\ \hline
    \end{tabular}
    \caption{Coupled signals from a scenario in which a detector main beam and a satellite main beam are pointed at each other, thus indicating the maximum coupled signal permissible by regulations. The band center and fractional bandwidth for the CMB bands in the first column can be found in Table \ref{tab:cmbs4}. The bandwidth assumed is the overlap between the CMB-S4 band and the bands for fixed satellite services in Table \ref{table:pfdlims}. The telescope forward gains are calculated for the 0.5-meter aperture SAT and the 5-meter aperture LAT. The resulting power can be compared to the typical power from the atmosphere, which is 1--5 pW for these detectors.}
    
    \label{table:mainmain}
    \end{table}

	In using these upper limits shown in Table \ref{table:pfdlims}, the highest possible transmitted flux densities under these regulations are compared with CMB-S4 instrument parameters to determine the most significant possible interference from operators in these bands.
	Power is assumed to be at the PFD limit, and the bandwidth used is the full overlap between the operator band and CMB instrument band.
	For a single-mode detector responding to both polarizations, the effective area-solid angle product is $\lambda^2$ for  observed wavelength $\lambda$.  Thus, $\lambda^2$ is the isotropic (0 dBi) cross-section, and multiplying $\lambda^2$ by the telescope forward gain factor gives the effective collecting area of the telescope when it is pointing at a satellite.  This collecting area multiplied by the satellite PFD limit corresponds to the case in which the satellite main beam and detector boresight are coincident.
    These calculations are enumerated in Table \ref{table:mainmain}, representing a worst-case scenario for power received by a single CMB detector.

\begin{figure}[ht]
	    \centering
	    \includegraphics[scale=0.5]{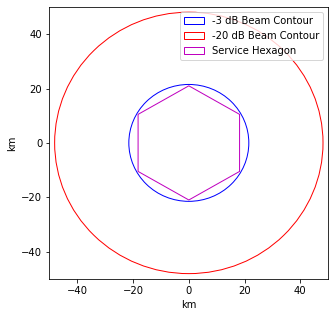}
	    \caption{Power reduction contours visualized as a function of distance from the beam center for a satellite at an altitude of 550\,km with a similar phased array to Starlink. \cite{slinkfiling2018} Understanding the beam strategies of satellites is imperative to modeling possible RFI impacts. The area of these service footprints is dependent upon the height of a satellite, its elevation in the sky, and its specific beam pattern.  }
	    \label{fig:contours}
	\end{figure}

	From here, reduction factors are added as they pertain to parameters of the operating constellation and the detector.
	For example, a reduction factor can be applied to account for scattered light events in which the satellite is not in the boresight of the telescope, but is still directing its main beam at it --- here, the small aperture telescopes may have a reduction of $-40$\,dB, and the large aperture telescopes may have a reduction of roughly $-20$\,dB, based on the shielding expectations described in Section \ref{ssec:sidelobes}. 
	The satellite may be within the main beam of the telescope, but it may be directing its beam far away from the observing location, such that the satellite signal experiences a reduction in PFD at the observing site based on the transmitter's sidelobe patterns. 
	A more common scenario would be the satellite passing through a far sidelobe while the satellite is directing its beam away from the observing location (sidelobe-sidelobe coupling), which may still result in a significant detection based on expected reduction factors.
	The specifics of this scenario depend upon the beam patterns of a particular satellite's design; for example, an early plan for Starlink (a SpaceX satellite-borne Internet service provider) involved satellites with phased arrays at 1150\,km orbits (now reduced to $\approx$ 550\,km orbits).
	These phased arrays would use synthesized beams which targeted a $-3$\,dB reduction at roughly 2.2\,degrees from the beam center, and a $-20$\,dB reduction at 5\,degrees.
	In this instance, this defines a serviceable hexagonal footprint within the half-power contour, as well as a $-20$\,dB reduction zone beyond that; this is visualized in Figure \ref{fig:contours}.
	The necessary exclusion region resulting on the ground would be over 100\,km in diameter, and detailed measurements would need to be made of the transmitter's sidelobe characteristics to determine reduction factors.
	This may be feasible for extremely remote locations, but would also require the cooperation of operators and potential customers across the proposed radio quiet zones.
	Another possibility would be the interruption of transmissions as satellites pass over radio quiet zones.
	While this could be extremely effective, the necessary exclusion zone for service becomes significantly larger.
	Under a simple model with satellites ceasing transmissions while higher than 45\,degrees above the horizon, the radius of the exclusion zone is roughly the satellite's orbital height.

	Another concern with satellite constellations with an impact that is complex to estimate is harmonic emission. 
	Harmonic emission is a specific type of spurious emission that occurs at integer multiples of the fundamental frequency.
	This unintentional harmonic emission generally results from nonlinearity and distortion within the transmitting electronics.
	Current regulatory recommendations pertaining to radio frequency emissions do consider emission beyond an allocated band. 
	Outside of allocated frequencies, the ITU has suggested a maximum power for spurious emission at a reduction of -60 dB from the power limit of said allocation.\cite{ITUspurems} 
	Harmonic emission can also behave in unpredictable ways relative to the fundamental frequency, which could reduce the directivity of the signal and eliminate reduction factors owing to the direction of the satellite signal. Thus, reduction factors here are reduced to telescope shielding alone. With the X-band allocation given in Table \ref{table:pfdlims}, harmonics could potentially affect every CMB band.
	Out-of-band emission can even impact protected radio astronomy bands. A prominent historical example was interference from the Iridium satellite constellation, which exceeded ITU-R Recommendation RA.769 and effectively had a signal 20 dB higher than the FCC regulation for a protected radio astronomy service band.\cite{Iridium_1996} Interference from this emission severely impacted radio astronomy imaging using the Very Large Array until filters were developed for mitigation. \cite{SpectMan_2009}
	
	
	
	
	The maximum allowed power for a direct coupling event between a communications satellite and a single CMB detector can be quite significant, as can be seen in Table \ref{table:mainmain}.
	The necessary reduction factors to mitigate the signal from a satellite passing above the horizon may not be achievable by CMB instrument design alone.
	The sensitivity of the CMB instruments is more complex than a single detector's coupling, and there is significant potential for insidious contamination from lower power events, as described further in Section \ref{sec:analysis_mitigation}.
	Testing and coordination will be more complex than the existing radio quiet zone agreements and the historical mitigations described in Section \ref{sec:rfi_pole}.
	Further study of satellite transmitters in cooperation with operators could provide better expectations on spurious emissions including harmonics, and beam patterns including far sidelobes. 
	Creating total exclusion zones around radio quiet zones to eliminate significant RFI from satellites is not straightforward, and would require coordination across a much broader geographical area than existing radio quiet zone agreements.
	Accurate orbital parameters (two-line elements, or TLEs), both real-time and archived, would help with forecasting satellite presence, as well as correlating satellite positions with observed signals in data, but are only a significant mitigation technique if interference from satellites remains relatively rare.

    \section{RFI Impacts on Observations and Techniques for Mitigation}\label{sec:analysis_mitigation}

    Section \ref{sec:coupling} described the instrument-level mitigation measures taken to prevent RFI from coupling to the instrument, but these are not 100\% effective, and do not mitigate coupling to most celestial signals such as satellite RFI as described in Section \ref{sec:satelliterfi}. In this section, we discuss the impact of RFI on observations, and give an overview of mitigation measures taken in analysis to remove spurious signals from the final survey, including maps and science results.
    The specific analysis methods used depend on details of the experiment including instrument design, survey strategy, and science goals. 
    For readers seeking a more comprehensive description of analysis methods, we include references throughout to relevant science papers for CMB experiments.
    We start with the impact on real-time data and a single observation in Section \ref{ssec:transients}. Years of observations are combined to create the survey data set as described in Section \ref{ssec:efficiency}.
    Maps of the sky at each frequency are used to construct a map of the cosmic microwave background, as described in Section \ref{ssec:componentseparation}.

    \subsection{Impact of Short Duration RFI Events on Survey Data}\label{ssec:transients}

    The study of the time-varying Universe is an increasingly active field, as instruments across the electromagnetic spectrum increase instantaneous sensitivity and sky coverage.  
    CMB instruments, although not designed for this purpose, are well suited for certain transient studies due to their sensitivity and observing strategy \cite{guns_2021,metzger2015, whitehorn2016, eftekhari2022}. 
    Real-time detection of astrophysical transients with CMB instruments is challenging for a variety of reasons, and RFI adds to these challenges.
    However, detection of astrophysical transients after some amount of post-processing and retrospective studies of transient phenomena in survey data are natural extensions of the techniques already used to precisely characterize the CMB.

    The real-time data from a CMB instrument is continuously-sampled power received by each individual detector. This time series is often referred to as time-ordered data (TOD).
    An individual detector's real-time data are dominated by non-astrophysical signals.
    Each detector has a power and noise level that is usually dominated by the residual atmosphere. 
    Typical ground-based instruments scan across the sky in azimuth while fixed at a constant elevation, keeping atmospheric power relatively constant.
    As the telescope scans across the sky, the signal recorded in each sweep is dominated by atmospheric brightness fluctuations from the residual atmosphere.
    By scanning relatively rapidly across the slow moving atmosphere, the structure of these fluctuations can be fully characterized for later removal (see for example, Ref. \citenum{morris2022}.
    In addition to atmospheric fluctuations, real-time data from the detectors also can contain glitches from short-lived RFI events.
    Glitches can also result from cosmic rays, as these energetic particles deposit power onto the TES bolometers\cite{Osherson2020}.
    A short-duration astrophysical event that is only seen by a single detector or a single pixel can be difficult to distinguish from these transient noise sources.
    Because the telescope is constantly moving, it is not straightforward to immediately identify and locate transient events from real-time detector data.
    At least one study has been done using archival time-ordered data from a CMB instrument to search for fast astrophysical transients, placing only an upper limit on possible fast radio bursts (FRBs) using SPTPol data\cite{harrington_2018}.

    The real-time data or recorded time-ordered data for an individual detector in a single scan are not particularly powerful because of the fundamental limitations described above.
    Rather, the sensitivity of CMB instruments comes from adding the data from many detectors, and adding these data over repeated identical scans.
    Combining data in this way not only increases the sensitivity of the resulting map, but is also a method of further rejecting atmospheric and other noise sources.
    Repeated identical scans typically occur over at least hour-long timescales, to build up significant statistics, and to allow the sky to drift relative to ground-fixed coordinates.
    Each scan's data will contain different transient signals, including the atmosphere and short-duration RFI.
    Sources that are fixed in ground coordinates (azimuth and elevation) can be distinguished from sources fixed in celestial coordinates (right ascension and declination), as they move relative to each other.

    Hourly and daily maps can be created through a computationally intensive map-making process, which combines the time-ordered data samples from each detector, filters out residual atmospheric fluctuations, and constructs a map in celestial coordinates. 
    For CMB analysis, analyzing the differences in these maps is done for data quality purposes, since time-varying phenomena are likely to be caused by instrumental effects or interference which must be understood and removed from the final cosmological data set. 
    Since the vast majority of expected RFI sources would not have a fixed location in celestial coordinates, they could be identified and removed from the maps at this stage so long as the emission is sufficiently long and bright to identify movement; very brief pulsed emission, including from sidelobes sweeping over the site from satellite movement, have been a contaminant in source searches.
    The same maps can also be studied for astrophysical transient phenomena.
    Recent examples of time-domain analysis include the detection of stellar flares \cite{naes_2020,guns_2021}, and constraints on time variations in the CMB from axions \cite{bk_axions_2022,Ferguson2022}.

    \subsection{Potential Impacts from RFI On Observing Efficiency and Survey Length}\label{ssec:efficiency}
    
    
    The scan strategy and map-making process described above allows for the identification and removal of certain kinds of RFI from the final cosmological data set.
    The removal of contaminated data can have a significant impact on the observing efficiency of the experiment, and the survey length necessary to achieve its science goals.
    The overall observing efficiency of a CMB experiment can be found after the survey is complete by comparing the final map depths to the instantaneous sensitivity and length of time that the survey ran.
    Due to a combination of weather conditions, instrument and detector down time, and stringent data quality cuts, the observing efficiency for ground-based CMB experiments has historically been around 25\%.
    Data cuts from RFI add another factor degrading observing efficiency that is outside of the experiment's control.
    Short duration, bright RFI events can be identified and removed at the detector or scan level, removing a small fraction of data if these events are rare.
    Historically, efforts have been made to track frequent bright RFI events to their source for mitigation or elimination, as described in Section \ref{sec:rfi_pole}.
    For a simple scenario of bright RFI events becoming much more frequent, the survey time must be extended proportional to the amount of data lost to RFI data cuts.
    Excision of a bright but short RFI event can also cause problems with reconstructing the signal at large angular scales, which are critical for CMB-S4's science goals.

    Not all RFI coupling is strong enough to immediately identify and remove.
    Insidious contamination can manifest only after significant amounts of data are added together.
    These types of systematic effects are typically discovered through statistical tests similar to the jacknife\cite{Tukey1958} and bootstrap methods\cite{Efron1979}.
    These analysis techniques check for statistical consistency across the survey's datasets, which are expected to be measuring the same astrophysical signal, the CMB\cite{bicep2_2014_ii, pb_bb_2020,spt_bb_2020,quiet_2012,planck_dataprocessing_2018}.
    Splitting the data along various axes including time and instrument configuration can reveal insidious contamination, and identify large sets of data that should be excluded.
    A simple example is the presence of the Sun or Moon in a far sidelobe causing contamination that requires certain data to be excluded from the cosmological analysis.
    An RFI-related example is the failure of these statistical tests on data taken during summer season at the South Pole, when this remote site has the highest level of human activity.
    Establishing a ``safe'' level of coupling for satellite RFI is especially difficult because of concerns about these subtle, insidious effects adding up over a years-long survey, and only becoming apparent relatively late in the cosmological analysis.
    Unlike obvious data cuts degrading sensitivity, the impact of this type of contamination would be hard to predict in advance, and possibly only apparent after the survey is complete.

    \subsection{Potential Impacts from RFI On CMB Maps and Cosmological Results}\label{ssec:componentseparation}

    As described in Section \ref{sec:intro}, CMB experiments must observe across a broad range of frequencies in order to characterize and remove astrophysical foregrounds that contaminate the cosmological signal.
    Each detector is sensitive to all incoming radiation across its frequency band, as defined by its on-chip band-defining filter, as described in Section \ref{ssec:opticalcoupling}.
    For CMB-S4, these bands are given by the band center and fractional bandwidth in Table \ref{tab:cmbs4}.
    Maps are made for a single frequency band, following the procedure of combining data from many detectors and many observations as described in Section \ref{ssec:efficiency}.
    These multi-frequency maps are used in a component separation analysis, where different sources of emission are distinguished using their unique frequency spectrum.
    The goal is a map of the cosmic microwave background, with other sources including dust and synchrotron emission removed.
    This process is not perfect, and the residual level of contamination from foregrounds remaining in the maps is estimated using simulations and statistical techniques.
    The broad frequency range of CMB-S4 and the distribution of detectors across frequencies are carefully designed to achieve a high level of foreground subtraction with very low residual contamination.
    The precision and accuracy of the final science results depends on the magnitude of this estimated residual contamination.
    
    With these broadband bolometer instruments, there is no ability to distinguish between frequencies within a band.
    If there is significant, long-term contamination from RFI within a certain frequency band for CMB-S4, the only option for excising it is to exclude all data taken by detectors of that frequency band.
    This is unlike channelized coherent instruments used in radio astronomy, which have some ability to do frequency excision for narrow channels within a band.
    To lose all data within a frequency band would have a significant impact on component separation analysis and the residual level of foreground contamination in the CMB map, which would then impact the scientific results.
    This could result from bright, continuous RFI from a source that cannot be mitigated or controlled.
    The frequency bands are set far in advance of observations and even production of the instrument, at the point where detector design is finalized.
    Because of this, reacting to unavoidable, significant contamination within a frequency band by shifting or narrowing a frequency band would be a relativey slow process.
    Recovering missing frequency information by designing a new frequency band would require a significant effort in re-designing and fabricating the detectors and supporting hardware.

    \section{Conclusions}
	
	By achieving order-of-magnitude increases in sensitivity and instantaneous sky coverage beyond that of current wide-band wide-field CMB experiments, CMB-S4 will reveal previously hidden features of the earliest stages of cosmological history while simultaneously greatly increasing our view of time-varying phenomena across the universe.  As we have discussed, observations over a wide range of frequency are required to disentangle the CMB from astrophysical foregrounds, and wide detection bandwidths are needed to achieve the required sensitivity.  These wide detection bandwidths necessarily encompass bands allocated for communications and other societally-important use.  CMB-S4 will build and improve upon existing CMB telescope and detector technology to minimize sensitivity to stray radiation and, like these earlier experiments, will be deployed to remote locations offering the best possible atmospheric conditions and freedom from terrestrial transmissions.  However, the rise of new telecommunications applications supported by dense LEO satellite constellations will require development of new modes of voluntary or regulatory coexistence between active services and passive scientific observers to ensure our continued view of the universe.

    \acknowledgments 
	
    CMB-S4 is supported by the Director, Office of Science, Office of High Energy Physics of the U.S. Department of Energy under Contract No.DE–AC02–05CH11231; by the National Energy Research Scientific Computing Center, a DOE Office of Science User Facility under the same contract; and by the Divisions of Physics and Astronomical Sciences and the Office of Polar Programs of the U.S. National Science Foundation under Mid-Scale Research Infrastructure award OPP-1935892. D.R.B. and I.E.B. are supported by the National Science Foundation's Division of Astronomical Sciences under award AST-2108704, and by the Department of Energy under the Office of High Energy Physics under award number DE-SC0021435. D.R.B was supported by the National Science Foundation's Office of Integrative Activities under award OIA-2033199. The authors thank Jeff Filippini and Bruce Partridge, members of the CMB-S4 collaboration, for their valuable feedback on this manuscript.

	\bibliography{rfi} 
	\bibliographystyle{spiebib} 
	
\end{document}